\documentclass[preprint,preprint,prc,showpacs,preprintnumbers,
                unsortedaddress,amsmath,amssymb,floatfix]{revtex4}

\usepackage{dcolumn}
\usepackage{bm}
\usepackage{longtable}

\usepackage{graphicx,epsfig,latexsym,amssymb}
\usepackage{multirow,amsmath,array,booktabs,color}
\usepackage[bookmarksnumbered,plainpages]{hyperref}
\usepackage[section]{placeins}

\newcommand{\beq}{\vspace{0.5em}\begin{equation}}
\newcommand{\eeq}{\end{equation}\vspace{0.5em}}
\newcommand{\beqn}{\vspace{0.5em}\begin{eqnarray}}
\newcommand{\eeqn}{\end{eqnarray}\par\vspace{0.5em}\noindent}

\newcommand{\beqa}{\vspace{0.5em}\begin{eqnarray*}}
\newcommand{\eeqa}{\end{eqnarray*}\par\vspace{0.5em}}
\newcommand{\bea}{\begin{array}}
\newcommand{\eea}{\end{array}}

\newcommand{\bcen}{\begin{center}}
\newcommand{\ecen}{\end{center}}
\newcommand{\btab}{\begin{tabular}}
\newcommand{\etab}{\end{tabular}}
\newcommand{\bsub}{\begin{subequations}}
\newcommand{\esub}{\end{subequations}}

\newcommand{\ovl}{\overline}
\newcommand{\unl}{\underline}

\newcommand{\balp}{\mbox{\boldmath$\alpha$}}

\newcommand{\bmu}{\mbox{\boldmath$\mu$}}

\newcommand{\brho}{\mbox{\boldmath$\rho$}}

\newcommand{\bsig}{\mbox{\boldmath$\sigma$}}

\newcommand{\bome}{\mbox{\boldmath$\omega$}}

\newcommand{\bSig}{\mbox{\boldmath$\Sigma$}}

\newcommand{\bj}{{\mathbf{j}}}

\newcommand{\bp}{{\mathbf{p}}}

\newcommand{\br}{{\mathbf{r}}}

\newcommand{\bA}{{\mathbf{A}}}
\newcommand{\bB}{{\mathbf{B}}}

\newcommand{\bL}{{\mathbf{L}}}

\newcommand{\bP}{{\mathbf{P}}}

\newcommand{\bS}{{\mathbf{S}}}

\newcommand{\bV}{{\mathbf{V}}}

\newcommand{\dfra}{\displaystyle\frac}

\begin{document}
\title{Time-odd triaxial relativistic mean field approach for
nuclear magnetic moment}
\author{J. M. Yao}
\affiliation{School of Physics, Peking University, Beijing 100871,
China}
\author{H. Chen}
\affiliation{School of Physical Science and Technology, Southwest
University, Chongqing 400715, China}
\author{J. Meng}
\email[E-mail:]{mengj@pku.edu.cn}
 \affiliation{School of Physics, Peking University, Beijing 100871, China}
 \affiliation{School of Physical Science and Technology, Southwest
University, Chongqing 400715, China}
 \affiliation{Institute of Theoretical Physics, Chinese Academy of
Sciences, Beijing 100080, China}
 \affiliation{Center of Theoretical Nuclear Physics, National
             Laboratory of Heavy Ion Accelerator, Lanzhou 730000, China}

\begin{abstract}
 \vspace{2em}

The time-odd triaxial relativistic mean field approach is developed
and applied to the investigation of the ground-state properties of
light odd-mass nuclei near the double-closed shells. The nuclear
magnetic moments including the isoscalar and isovector ones are
calculated and good agreement with Schmidt values is obtained.
Taking $^{17}$F as an example, the splitting of the single particle
levels ( around $~0.7$ MeV near the Fermi level ), the nuclear
current, the core polarizations, and the nuclear magnetic potential,
i.e., the spatial part of the vector potential, due to the violation
of the time reversal invariance are investigated in detail.

\end{abstract}
 \maketitle

%
 \section{Introduction}

Nuclear magnetic moment is one of the most important physics
observables. It provides a highly sensitive probe of the single
particle structure, serves as a stringent test of the nuclear
models, and has attracted the attention of nuclear physicists since
the early days\cite{Stoyle56,Wildenthal79,Arima84}. Up to now, the
static magnetic dipole moments of ground states and excited states
of atomic nuclei throughout the periodic table have already been
measured with several methods \cite{Stone01}. With the development
of the radioactive ion beams (RIB) technique, it is now even
possible to measure the nuclear magnetic moments of many short-lived
nuclei near the proton and neutron drip lines with high precision
\cite{Ueno96}.

Lots of efforts have been made to describe the nuclear magnetic
moment non-relativistically or relativistically. In the
non-relativistic theory, it has been pointed out early that the
single-particle state of the shell model can couple to more
complicated 2p-1h configurations\cite{Arima54} and there are mesons
exchange corrections caused by the nuclear medium effect
\cite{Chemtob69}. With modified g-factor or configuration mixing
effect, non-relativistic approaches have turned out to be successful
in reproducing data \cite{Wildenthal79}. Therefore the single
particle picture in the mean field approach may not be expected to
describe the magnetic moment well\cite{Arima87}. For LS closed shell
nuclei plus or minus one nucleon, however, there is no spin-orbit
partners on both sides of the Fermi surface and the magnetic moment
operator can not couple to magnetic resonance \cite{Hofmann88}.
Furthermore, the contributions from the pion-exchange current as
well as others processes, e.g. $\Delta$-hole excitation, to
iso-scalar current have turned to be very small in the first order
approximation \cite{Towner83,Ichii87}. Therefore, the iso-scalar
magnetic moments of the light nuclei near double-closed shells
should be described well in single-particle picture.

In relativistic approach, the single particle wave function of
nucleon is described by Dirac spinor with the large and small
components, resulting in the relativistic effect in nuclear current
and magnetic moment. Furthermore, as the spin-orbit coupling appears
naturally in the relativistic approach, it is expected that the
anomalous magnetic moment, which is related to the spin of nucleons,
can be described well in relativistic approach with the free
g-factor for the nucleon.

During the last two decades relativistic mean field (RMF) theory
has achieved great
success~\cite{Serot86,Reinhard89,Ring96,meng05ppnp} in describing
many nuclear phenomena for stable nuclei, exotic nuclei
\cite{meng96prl,meng98prl}, as well as supernova and neutron stars
\cite{Glendenning}. The RMF theory incorporates many important
relativistic effects from the beginning, such as a new saturation
mechanism by the relativistic quenching of the attractive scalar
field, the existence of anti-particle solutions, the Lorentz
covariance and special relativity which make the RMF theory more
appealing for studies of high-density and high temperature nuclear
matter, the origin of the pseudospin symmetry
\cite{Arima69,Hecht69} as a relativistic symmetry
\cite{Ginocchio97,meng98prc,meng99prc}, and spin symmetry in the
anti-nucleon spectrum\cite{zhou03prl}, etc. Recently, good
agreements with existing data for the ground-state properties of
over 7000 nuclei has been achieved in the RMF+BCS model
\cite{Geng05PTP}.

However, it was found out that a straightforward application of the
single-particle relativistic model, where only sigma and the
time-component of the vector mesons were considered, can not
reproduce the experimental magnetic
moments~\cite{Ohtsubo73,Miller75,Bawin83,Bouyssy84,Kurasawa85}. It
is attributed to the small renormalized nucleon mass ($M^*\sim
0.6M$) which results in the enhancement of the relativistic effect
on the Dirac current\cite{Shepard88}. As an improvement, the vertex
corrections have to be introduced to define effective
single-particle currents in nuclei, e.g., the "back-flow" effect in
the framework of the relativistic extension of Landau's Fermi-liquid
theory~\cite{McNeil86} or the RPA type summation of p-h and
p-$\bar{n}$ bubbles in relativistic Hartree
approximation~\cite{Ichii87,Shepard88}.

In the widely used RMF approaches, there are only the time-even
fields which are essential to physical observable. In odd-A or
odd-odd nuclei, however, the Dirac current due to the unpaired
valence nucleon will lead to the time-odd component of vector
fields, i.e., the nuclear magnetic potential. In other words, the
time-odd fields will have the same effect as the "back-flow" effect
or the RPA type summation of p-h and p-$\bar{n}$ bubbles and give
rise to the core polarization which will modify the nuclear current,
single-particle spin and angular momentum, giving the appropriate
magnetic moments. In fact, the time-odd fields are very important
for the description of the magnetic moments
\cite{Hofmann88,Furnstahl89}, and rotating nuclei
\cite{Konig93,Mad00}, etc.

It is thus essential to consider the spatial-component of vector
fields in the framework of RMF theory to investigate the
single-particle properties in odd-mass nuclei. With the time-odd
nuclear magnetic potential in axial deformed RMF, the iso-scalar
magnetic moment have been reproduced
well~\cite{Hofmann88,Furnstahl89}. Therefore, it is interesting to
include the time-odd potential into triaxial case \cite{Meng06PRC}
and investigate the distribution of non-vanishing spatial-component
of the $\omega$ vector meson field, nuclear current, magnetic
potential and magnetic moments in nuclei with odd numbers of protons
and/or neutrons. Taking $^{17}$F as an example, the density
distribution, single-particle energy and its splitting due to the
violation of the time reversal invariance, nuclear potential,
orbital momentum, etc., will be investigated. The importance of the
time-odd component in the description of the nuclear current and
magnetic moment will be illustrated.

The paper is arranged as the following. In Sec.\ref{sec:sec1}, the
time-odd triaxial RMF approach is presented in detail. The splitting
due to the violation of the time reversal invariance in the
single-particle energy will be estimated by reducing the Dirac
equation with the time-odd nuclear magnetic potential to
Schr\"{o}dinger equation. The numerical details for solving the
triaxial RMF equations expanded in three dimensional harmonic
oscillator basis are given in Sec.\ref{sec:sec2}. The results and
discussions are presented in Sec.\ref{sec:sec3}, and a short summary
is given in Sec.\ref{sec:sec4}.


 \section{Time-odd triaxial relativistic mean field approach}
 \label{sec:sec1}


 The starting point of the RMF theory is the standard effective Lagrangian density
constructed with the degrees of freedom associated with the
nucleon field($\psi$), two isoscalar meson fields ($\sigma$ and
$\omega$), the isovector meson field ($\rho$) and the photon field
($A$)~\cite{Serot86,Reinhard89,Ring96,meng05ppnp}:
\begin{widetext}
 \begin{eqnarray}
  \label{lagrangian}
   \displaystyle
   {\cal L}
     & = &
      \bar\psi\left[i\gamma^\mu\partial_\mu - M
            - g_\sigma\sigma
            - g_\omega\gamma^\mu\omega_\mu
            - g_\rho\gamma^\mu\vec\tau\cdot\vec\rho_\mu
            - e \gamma^\mu A_\mu\frac{1-\tau_3}{2}\right]\psi \nonumber\\
     &   & \mbox{}
       + \frac{1}{2} \partial_\mu \sigma \partial^\mu \sigma
       - \frac{1}{2} m_\sigma^2\sigma^2-\frac{1}{3}g_2\sigma^3-\frac{1}{4}g_3\sigma^4
    \nonumber \\
   &   & \mbox{}
    - \frac{1}{4} \Omega_{\mu\nu} \Omega^{\mu\nu}
    + \frac{1}{2} m_\omega^2 \omega_\mu \omega^\mu+\frac{1}{4}c_3(\omega_\mu \omega^\mu)^2
   \nonumber \\
   &   & \mbox{}
    - \frac{1}{4} \vec R_{\mu\nu} \vec R^{\mu\nu}
    + \frac{1}{2} m_{\rho}^{2} {\vec \rho}_\mu\cdot{\vec\rho}^\mu
   \nonumber \\
   &   & \mbox{}
    - \frac{1}{4} F_{\mu\nu} F^{\mu\nu},
\end{eqnarray}
\end{widetext}
in which the field tensors for the vector mesons and the photon
are respectively defined as,
 \begin{eqnarray}
 \displaystyle
 \left\{
  \begin{array}{rcl}
   \Omega_{\mu\nu}  & = & \partial_\mu\omega_\nu
                         -\partial_\nu\omega_\mu, \\
   \vec R_{\mu\nu} & = & \partial_\mu\vec \rho_\nu
                         -\partial_\nu\vec \rho_\mu, \\
   F_{\mu\nu}       & = & \partial_\mu {A}_\nu
                         - \partial_\nu {A}_\mu.
  \end{array}
  \right.
  \label{eq:tensors}
 \end{eqnarray}

 The Lagrangian (\ref{lagrangian}) includes respectively the
 non-linear self-coupling of the $\sigma$-meson and the
 $\omega$-meson. In this paper the arrows are
 used to indicate vectors in isospin space and the bold types for
 the space vectors. The parameters include the effective meson
 masses, the meson-nucleon couplings and the
 non-linear self-coupling of the mesons. Because of charge conservation,
 only the 3rd-component of the iso-vector $\rho$-meson contributes
 (for brevity the superscript $'3'$ is dropped out in the following).

 From the Lagrangian in Eq.(\ref{lagrangian}), the equations of motion for
 nucleon is
 \beq
  \label{DiracEq}
  \{\balp\cdot[-i\nabla-\bV(\br)]+\beta M^*+V_0(\br)\}\psi_i(\br)=\epsilon_i\psi_i(\br),
 \eeq
where $M^*(\br)$ is defined as
 \beq
 M^*(\br)=M+S(\br),
 \eeq
 with the attractive scalar potential S($\br$),
 \beq
 \label{scapot}
  S(\br)=g_\sigma\sigma(\br),
 \eeq
 and the usual repulsive vector potential, i.e., the time component of the vector
 potential, $V_0(\br)$,
 \beq
  \label{vecpot}
  V_0(\br)=g_\omega\omega_0(\br)+g_\rho\tau_3\rho_0(\br)
  +e\frac{1-\tau_3}{2}A_0(\br).
 \eeq
 The time-odd nuclear magnetic potential, i.e., the spatial
 components of the vector potential, $\bV(\br)$,
 \beq
  \label{magpot}
  \bV(\br)=g_\omega\bome(\br)+g_\rho\tau_3\brho(\br)+e\frac{1-\tau_3}{2}\bA(\br).
 \eeq
is due to the spatial component of the vector fields. Compared with
$\bome(\br)$ field, $\brho(\br)$ and $\bA(\br)$ fields turned out to
be small in  $\bV(\br)$ so that they were often neglected for light
nuclei\cite{Hofmann88}.

The Klein-Gordon equations for the mesons and the electromagnetic
fields are
 \beq
  \label{MesonEqs}
  (-\nabla^2+m^2_\zeta)\zeta(\br)=S_\zeta(\br),
 \eeq
where,
 \beqn
  \label{Sources}
   S_\zeta(\br)
         = \left\{
           \begin{array}{ll}
           \vspace{0.3cm}
             -g_\sigma\rho_s(\br)-g_2\sigma^2(\br)-g_3\sigma^3(\br), & \zeta=\sigma  \\
             \vspace{0.3cm}
             g_\omega j^\mu_B(\br)-c_3\omega^\nu\omega_\nu\omega^\mu(\br), &\zeta=\omega^\mu\\
             \vspace{0.3cm}
             g_\rho {j}_R^\mu(\br)                                   , &\zeta=\rho^\mu_3\\
             ej^\mu_c(\br)                                        ,&   \zeta=A^\mu
           \end{array}
          \right.
 \eeqn
with,
 \beqn
  \label{Densities}
  \displaystyle
  \left.
   \begin{array}{lcl}
     \rho_s(\br) &=& \sum\limits_{i}n_i\bar\psi_i(\br)\psi_i(\br),\\
     j^\mu_B(\br) &=& \sum\limits_{i}n_i\bar\psi_i(\br)\gamma^\mu\psi_i(\br),\\
     {j}_R^\mu(\br) &=& \sum\limits_{i}n_i\bar\psi_i(\br)\gamma^\mu\tau_{i3}\psi_i(\br),\\
     j^\mu_c(\br) &=&
     \sum\limits_{i}n_i\bar\psi_i(\br)\gamma^\mu\dfra{1-\tau_{i3}}{2}\psi_i(\br).
    \end{array}
  \right.
 \eeqn
 The sums are taken over the particle states only, i.e. the
contributions from the negative-energy states are neglected
(no-sea approximation), and $n_i$ is the occupancy probability of
state $\psi_i(\br)$, which satisfies the normalization condition
 \beq
  \int d^3r \sum_i n_i\psi_i^\dagger(\br)\psi_i(\br)={\rm Z}~~~ ({\rm or}~{\rm N}),
 \eeq
where Z (N) is the proton (neutron) number. If the pair correlation
is neglected, the occupancy $n_i$ is one or zero for the states
below or above the Fermi surface.

 The total energy of the system, including the spatial-component of vector fields, is,
 \beq
  E_{\rm total}=E_{\rm part.}+E_\sigma+E_\omega+E_\rho+E_\gamma+E_{\rm cm}+E_{\rm
  pair},
 \eeq
 where,
 \begin{widetext}
 \beqn
  \begin{array}{lcl}     \vspace{0.2cm}
   E_{\rm part.}&=&\displaystyle \sum^A\limits_{i=1}n_iE_i,\\
   \vspace{0.2cm}
   E_\sigma     &=&-\dfra{1}{2}\int d^3r\left[g_\sigma\rho_s(\br)\sigma(\br)
                    +\dfra{1}{3}g_2\sigma^3(\br) +\dfra{1}{2}g_3\sigma^4(\br)\right],\\
   \vspace{0.2cm}
   E_\omega     &=&-\dfra{1}{2}\int d^3r\left[g_\omega \omega_0(\br) j^0_B(\br)
             -g_\omega \bome(\br)\cdot{\bj}_B(\br)
             -\dfra{1}{2}c_3(\omega_\mu \omega^\mu)^2(\br)\right],\\
    \vspace{0.2cm}
   E_\rho   &=&-\dfra{1}{2}\int d^3r g_\rho\rho_0(\br) j^0_R(\br),\\
   E_\gamma &=&-\dfra{e^2}{8\pi}\int d^3r A_0(\br) j^0_c(\br),
 \end{array}
 \label{Eq:MesonEngrgy}
 \eeqn
\end{widetext}
 with $E_i=\epsilon_i-M$ and $E_{\rm pair}$ is due to the pairing correlation.

 The center-of-mass correction to the binding energy from projection-after-variation
 in first-order approximation is given by~\cite{Bender00},
  \beq
   \label{Cent}
   E_{\rm cm}  = -\dfra {1}{2mA}\langle \hat\bP^2_{cm}\rangle.
  \eeq
The root-mean-square ($rms$) radius is defined as,
  \beqn
  \displaystyle
     \langle r^2\rangle^{1/2} =\left[\dfra {\displaystyle\int d^3r r^2\rho(\br)}
                                  {\displaystyle\int d^3r
                                  \rho(\br)}\right]^{1/2}.
   \eeqn
 The quadrupole moments $Q^{n,p}_{20}$ and $Q^{n,p}_{22}$  for neutron and proton are
 respectively calculated by
 \beqn
   Q^{n,p}_{20} =\sqrt{\frac{5}{16\pi}}\langle
   2x^2-y^2-z^2\rangle,\nonumber\\
   \quad
      Q^{n,p}_{22} =\sqrt{\frac{15}{32\pi}}\langle z^2-y^2\rangle.
   \eeqn
 The deformation parameters $\beta$ and $\gamma$ can
be obtained respectively by the corresponding quadrupole moments
for neutron, proton and matter (e.g., $Q_{20} =
Q^{n}_{20}+Q^{p}_{20}$) as,
 \beq
  \displaystyle
  \beta=\frac{4\pi}{3A R^2_0}
  \sqrt{Q^2_{20}+2Q^2_{22}},
   \quad
  \gamma \simeq \tan^{-1}(\sqrt{2}\frac{Q_{22}}{Q_{20}}),
 \eeq
with $Q_0 = \displaystyle\sqrt
 {\frac{16\pi}{5}}\sqrt{Q^2_{20}+2Q^2_{22}}$ and
$R_0=1.2A^{1/3}$ fm.


 Before any sophisticated numerical calculation, it is interesting
 to investigate the time-odd potential in the Dirac equation
 qualitatively by reducing the
 Dirac equation to more familiar Schr${\rm\ddot{o}}$dinger-like form
 and estimate the splitting of the time-reversal orbits due to the
 time-odd potential $g_\omega\bome$.
 The Dirac equation
Eq.(\ref{DiracEq}) can be reduced to the
Schr${\rm\ddot{o}}$dinger-like equation for the large component
as,
 \begin{widetext}
  \beq
   \label{Schrodinger}
   [\bsig\cdot(\bp-g_\omega\bome)\frac{1}{\epsilon+M+S-V_0}\bsig\cdot(\bp-g_\omega\bome)+V_0+S]f
   =(\epsilon-M)f.
   \eeq
  \end{widetext}
The time-odd part of the Hamiltonian can be obtained as the
following:
 \begin{widetext}
  \beq
    H_{\mbox{odd}} = -\frac {g_\omega}  {E+2M+S-V_0}
                     (\bsig\cdot\bome)  (\bsig\cdot\bp)
                    -\bsig\cdot\bp
                     \frac  {g_\omega}{E+2M+S-V_0}
                     \bsig\cdot\bome,
   \eeq
  \end{widetext}
 Introducing  $\displaystyle\bome=\frac{1}{2}\bB_n\times\br$, which
 defines a nuclear magnetic field $\bB_n = \nabla \times \bome$, the time-odd
 part of the Hamiltonian can be rewritten as:
   \beqn
    \displaystyle
     H_{\mbox{odd}} = -\frac {g_\omega} {E+2M+S-V_0}
                       (\bL+2\bS)\cdot\bB_n,
   \eeqn
 where the assumption $\bome \times \bp =0$ is used and the higher
 order term $\displaystyle, \sim \frac{1} {(E+2M+S-V_0)^2}$, is
 neglected.

 It indicates that the spatial-component of vector meson field $\bome$
 will lead to a time-odd term in the Schr${\rm \ddot{o}}$dinger-like
 equation and result in the energy splitting.
 The Kramer's degeneracy in odd-mass nuclei is thus removed
 by the effective intrinsic nuclear
  magnetic dipole moment,
 ${\displaystyle \frac{-g_\omega}{E+2M+S-V_0}(\bL+2\bS)}$,
 coupled to the nuclear magnetic field $\bB_n$. The magnetic field
 $\bB_n$, as shown in the following, is around
 $\bB_n\sim5\times 10^{-4}$fm$^{-2}$ in $^{17}$F,
 which will lead to the energy splitting between two time reversal conjugate states,
 e.g., $\sim$0.4 MeV for $1s_{1/2}$. Of course, the more realistic splitting
 will depend on the detailed information on the nuclear magnetic field
 $\bB_n$, the scalar and vector potentials and the density
 distribution of the particular orbits, which can be obtained by solving the
 Dirac equation (\ref{DiracEq}) self-consistently.

\section{Numerical Details}
 \label{sec:sec2}


 For the triaxial deformed nucleus, the Dirac spinor can be expanded
 by using the three-dimensional harmonic
 oscillator wave function  $\Phi_{\unl\alpha}(\br,s)$ and
$\Phi_{\tilde{\bar{\alpha}}}(\br,s)$ in Cartesian coordinates:
 \begin{widetext}
  \beqn
    \displaystyle
  \label{Basis}
   \left\{
    \begin{array}{lclcl}
     \Phi_{\unl\alpha}(\br,s)&=&
      \phi_{n_x}(x)\phi_{n_y}(y)\phi_{n_z}(z)
      \displaystyle \frac{i^{n_y}} {\sqrt{2}}
       \left(
        \begin{array}{c}
        1\\
        (-1)^{n_x+1}
        \end{array}
       \right) & {\rm with} & \unl\alpha=\vert n_x, n_y, n_z, m_s\rangle,   \\
      \Phi_{\tilde{\bar{\alpha}}}(\br,s)&=&
     \phi_{\tilde n_x}(x)\phi_{\tilde n_y}(y)
     \phi_{\tilde n_z}(z)\displaystyle\frac{i^{\tilde n_y}}{\sqrt{2}}(-1)^{\tilde n_x+\tilde n_y+1}
      \left(
       \begin{array}{c}
       1\\
       (-1)^{\tilde n_x}
       \end{array}
      \right) & {\rm with} & {\bar{\alpha}}=\vert \tilde n_x, \tilde n_y, \tilde n_z, \tilde m_s\rangle,
   \end{array}
  \right.
 \eeqn
 \end{widetext}
where, the phase factor $i^{n_y}$ is chosen in order to have a real
matrix elements for the Dirac equation \cite{Girod83}. The time
reversal conjugate states $\Phi_{\unl\alpha}(\br,s)$ and
$\Phi_{\tilde{\bar{\alpha}}}(\br,s)$ satisfy,
 \beq
  \hat T\Phi_{\unl\alpha}(\br,s)=-\Phi_{\tilde{\bar{\alpha}}}(\br,s),~~~
  \hat T\Phi_{\tilde{\bar{\alpha}}}(\br,s)=\Phi_{\unl\alpha}(\br,s).
 \eeq
with $\hat T\equiv-i\sigma_y\hat K$.

 The normalized oscillator function $\phi_{n_k}(k)$ in the $k$ direction
 ($k$ denotes x, y, or  z) is,
 \beqn
  \displaystyle
  \phi_{n_k}(k)=\frac{N_{n_k}}{\sqrt{b_k}}H_{n_k}(\frac{k}{b_k})
  \exp [{- \dfra 1 2 (\dfra {k}{b_k})^2}],
 \eeqn
with the normalization factor
$N_{n_k}=(\sqrt{\pi}2^{n_k}n_k!)^{-1/2}$. The Hermite polynomial
$H_{n}(\xi)$ is given by
 \beq
  H_{n}(\xi)=(-1)^{n}e^{\xi^2}\frac{d^n}{d\xi^n}e^{-\xi^2}.
 \eeq
 The frequency in
$k$-direction $\omega_k$ can be written in terms of the deformation
parameters $\beta$ and $\gamma$ as,
 \beq
  \hbar\omega_k=\hbar\omega_0\exp[\sqrt{\frac{5}{4\pi}}\beta\cos(\gamma-\frac{2k\pi}{3})],
 \eeq
with oscillator length $\displaystyle b_k=\sqrt{ {\hbar} /
{M\omega_k}}$ and $M$ the mass of the nucleon. The oscillator
frequency is $\hbar\omega_0 = 41A^{-1/3}$ MeV and the corresponding
oscillator length $\displaystyle b_0=\sqrt{ {\hbar} / {M\omega_0}}$.

 The Dirac spinor for nucleon has the form
 \beqn
  \label{WaveFunction}
   \psi_i(\br)=
    \left(
     \begin{array}{l}
     f(\br,s)\\
     ig(\br,s)
    \end{array}
   \right)\chi_{i}(t),
 \eeqn
where $\chi_{i}(t)$ is the isospin part. In odd-A nuclei, as the
time reversal invariance is broken by the unpaired valence nucleon,
Eq.(\ref{WaveFunction}) can be written as the linear combination of
time reversal conjugate basis~\cite{Koepf89},
 \beqn
 \label{combination}
 \left\{\begin{array}{ccc}
  f(\br,s)&=&\sum\limits_{\unl\alpha}f_{\unl\alpha}\vert \unl\alpha\rangle
            + \sum\limits_{\ovl {\alpha}}f_{\ovl {\alpha}}\vert\ovl {\alpha}\rangle,\\
  g(\br,s)&=&\sum\limits_{\unl\alpha}g_{\unl\alpha}\vert \unl\alpha\rangle
          + \sum\limits_{\ovl {\alpha}}g_{\ovl {\alpha}}\vert\ovl {\alpha}\rangle.
  \end{array}
   \right.
 \eeqn
However, one can define the time reversal conjugate states
${\psi_{\unl j}} = \hat T \psi_{\ovl j}$ and $\psi_{\ovl j} = -\hat
T \psi_{\unl j}$ as,
  \beqn
  \label{TimeOdd}
   \psi_{\unl j}(\br,t,s)=
    \left(
     \begin{array}{l}
      \sum\limits_{\unl\alpha}f_{\unl\alpha}^{\unl j}\Phi_{\unl\alpha}(\br,s)\\
      \sum\limits_{\tilde{\bar\alpha}}ig_{\tilde{\bar\alpha}}^{\unl j}\Phi_{\tilde{\bar\alpha}}(\br,s)
    \end{array}
   \right)\chi_{t_{\unl j}}(t),\nonumber\\
  \label{TimeEven}
  \psi_{\ovl j}(\br,t,s)=
   \left(
    \begin{array}{l}
     \sum\limits_{\tilde{\bar\alpha}}f_{\tilde{\bar\alpha}}^{\ovl j}\Phi_{\tilde{\bar\alpha}}(\br,s)\\
     \sum\limits_{\unl\alpha}ig_{\unl\alpha}^{\ovl j}\Phi_{\unl\alpha}(\br,s)\\
    \end{array}
    \right)\chi_{t_{\ovl j}}(t),
  \eeqn
 so that the Dirac equation for the nucleons can be solved separately
 in subspaces $\{\psi_{\unl j}\}$ or $\{\psi_{\ovl j}\}$, i.e.,
 \beqn
 \displaystyle
 \label{Matrix1}
  \left(
   \begin{array}{cc}
   {\cal\unl A_{\alpha'\alpha}} & {\cal\unl B_{\alpha'\tilde\alpha}}\\
   {-\cal\unl B}_{\tilde{\alpha}'\alpha} & {\cal\unl C_{\tilde{\alpha}'\tilde\alpha}}
    \end{array}
   \right)
   \left(
     \begin{array}{l}
      f^{\unl j}_{\alpha}\\
      g_{\tilde\alpha}^{\unl j}
    \end{array}
   \right)=\varepsilon_{\unl j}
   \left(
     \begin{array}{l}
      f^{\unl j}_{\alpha^\prime}\\
      g_{\tilde\alpha^\prime}^{\unl j}
    \end{array}
   \right),
 \eeqn
 and
 \beqn
  \label{Matrix2}
  \left(
   \begin{array}{cc}
   {\cal\ovl A_{{\tilde\alpha}'{\tilde\alpha}}} &{\cal\ovl B_{{\tilde\alpha}'\alpha}}\\
   {-\cal\ovl B}_{\alpha'{\tilde\alpha}}& {\cal\ovl C_{\alpha'\alpha}}
    \end{array}
   \right)
   \left(
    \begin{array}{l}
     f^{\ovl j}_{\tilde\alpha}\\
    g_{\alpha}^{\ovl j}\\
    \end{array}
    \right)=\varepsilon_{\ovl j}
    \left(
    \begin{array}{l}
     f^{\ovl j}_{\tilde\alpha^\prime}\\
    g_{\alpha^\prime}^{\ovl j}\\
    \end{array}
    \right).
 \eeqn
 More details can be found in
 Appendix.

 Using the relations $\hat T\bsig\hat T^{-1}=-\bsig$ and $\hat T\bV \hat T^{-1}=
 \bV$, one can have
 \beq
   \langle \unl\alpha \vert\bsig\cdot i\bV(\br)\vert \ovl\alpha\rangle
   =-\langle \ovl\alpha \vert \bsig\cdot i\bV(\br)\vert \unl\alpha\rangle,
 \eeq
 which will lead to different Hamiltonian matrix elements in subspaces
 $\{\psi_{\unl j}\}$ and $\{\psi_{\ovl j}\}$, as shown in
 Eqs. (\ref{Matrix1}) and (\ref{Matrix2}). Therefore
 the time-odd nuclear magnetic potential in Eq.(\ref{magpot})
 will result in the violation of the time reversal invariance
 and the splitting
 between the time reversal conjugate states
 $\varepsilon_{\unl j}$ and $\varepsilon_{\ovl j}$.

The meson fields $\zeta(\br)$ are also expanded in three dimensional
harmonic oscillator basis with the same deformation parameters
$\beta$ and $\gamma$,
 \beq
  \zeta(\br)=\sum_{n_x, n_y, n_z}\zeta_{n_x, n_y, n_z}\phi_{n_x}(x)\phi_{n_y}(y)\phi_{n_z}(z).
 \eeq
In order to simplify the calculations and to avoid additional
parameters, the oscillator length for mesons $b_B$ is smaller by a
factor of $\sqrt{2}$ than that $b_0$ for the nucleons. The
Klein-Gordon equations (\ref{MesonEqs}) for the meson fields become
the inhomogeneous sets of linear equations,
 \begin{widetext}
 \beqn
  \label{gordenequation}
  \sum_{n_x n_y n_z}^{N_B} {\cal M}_{n^{''}_x n^{''}_y n^{''}_z,n_x n_y n_z}\zeta_{n_x, n_y, n_z}
    = \int dxdydz\phi_{n^{''}_x}(x)\phi_{n^{''}_y}(y)\phi_{n^{''}_z}(z)S_\zeta(x,y,z),
 \eeqn
 \end{widetext}
 with the matrix elements,
 \begin{widetext}
 \beqn
  \label{omegamatrix}
  {\cal M}_{n^{''}_x n^{''}_y n^{''}_z,n_x n_y n_z}
         &=&-\frac{1}{b^2_x}[\sqrt{n_x(n_x-1)}\delta_{n^{''}_x n_x-2}+\sqrt{(n_x+1)(n_x+2)}\delta_{n^{''}_x n_x+2}]\delta_{n^{''}_y n_y}\delta_{n^{''}_z n_z}\nonumber\\
         &&-\frac{1}{b^2_y}[\sqrt{n_y(n_y-1)}\delta_{n^{''}_y n_y-2}+\sqrt{(n_y+1)(n_y+2)}\delta_{n^{''}_y n_y+2}]\delta_{n^{''}_x n_x}\delta_{n^{''}_z n_z}\nonumber\\
         &&-\frac{1}{b^2_z}[\sqrt{n_z(n_z-1)}\delta_{n^{''}_z n_z-2}+\sqrt{(n_z+1)(n_z+2)}\delta_{n^{''}_z n_z+2}]\delta_{n^{''}_x n_x}\delta_{n^{''}_y n_y}\nonumber\\
         &&+2[\frac{1}{b^2_x}(n_x+\frac{1}{2})+\frac{1}{b^2_y}(n_y+\frac{1}{2})+\frac{1}{b^2_z}(n_z+\frac{1}{2})+\frac{1}{2}m^2_\omega]
         \delta_{n^{''}_x n_x}\delta_{n^{''}_y n_y}\delta_{n^{''}_z n_z}\nonumber\\
 \eeqn
 \end{widetext}
The source terms in Eq.(\ref{Densities}) can be obtained from the
nucleon wave functions in two subspaces. For the Coulomb field, due
to its long range character, the standard Green function method is
used\cite{Vautherin73}.

The expansion for the spinors and meson fields has to be truncated
to a fixed number of basis states and the cut-off is chosen in such
a way that the convergence is achieved for the binding energies as
well as the deformation parameters $\beta$ and $\gamma$. Here in the
calculation, up to $n_{f}=12$ and $n_{b}=10$ are adopted, which
gives an error less than 0.1\% for the binding energy as
demonstrated in Ref. \cite{Meng06PRC}, and the effective interaction
parameter sets PK1~\cite{Long04} has been used. As shown already in
Ref.\cite{Meng06PRC}, the binding energy and the deformations of the
nucleus are independent on the deformation of the basis in Eq. (25).
Therefore for convenience, the deformation of the basis in Eq. (25)
are chosen as $\beta=0$, i.e. , $\displaystyle b_k=\displaystyle
b_0$.  The center-of-mass correction is considered microscopically
as in Eq.(\ref{Cent}). The expanded oscillator shells for the small
component $g$ is carried out with $n_g = n_f+1$. In the calculation,
the mirror symmetry with respect to the $xy$, $xz$, and $yz$ planes
for the densities and rotational symmetry for the currents are
assumed to avoid the complex coefficients in the basis, which is
appropriate as illustrated in Ref.~\cite{Konig93,Meng06PRC}.

 \section{Results and Discussion}
 \label{sec:sec3}

Using Woods-Saxon potentials as initial potentials in
Eqs.(\ref{scapot}) and (\ref{vecpot}) in the Dirac
equation(\ref{DiracEq}), and one tenth of the time-component of the
vector meson field as the initial spatial-component, the above Dirac
equation (\ref{DiracEq}) and Klein-Gordon equations (\ref{MesonEqs})
are solved self-consistently by iteration.

Taking $^{17}$F as an example, the importance for taking into
account self-consistently the spatial-component of the $\omega$
field due to the violation of time reversal symmetry in odd-mass
nuclei has been demonstrated by examining the nuclear magnetic
potential, the nuclear magnetic field, the ground-state properties
including the binding energy, $rms$ radii and deformation $\beta$
and $\gamma$, the single-particle energy and the splitting of the
time reversal conjugate states, the density and current
distribution, etc.

Finally, the nuclear magnetic moments for the light double-closed
shells nuclei plus or minus one nucleon, including the Dirac,
anomalous, iso-scalar, and iso-vector magnetic moments, will be
investigated and compared with data and former calculation results
available.

%
%
 \subsection{Nuclear magnetic potential}
 \label{sec:NM Potential}
%
%
%
\begin{figure}[ht]
 \centering
  \includegraphics[width=8cm]{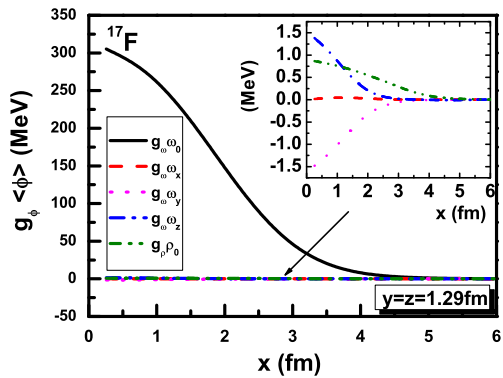}
   \caption{(color online) The distribution of the time-component of the
    vector potential and the nuclear magnetic potential, i.e., the spatial
    components of the vector potential, versus $x$ axis for $y=z=1.29$ fm
    in $^{17}$F.}
   \label{fig1}
 \end{figure}
%
There is no time-odd magnetic potential in the ground-state of
even-even nucleus because of the time-reversal symmetry. For odd-odd
or odd A nucleus, the unpaired valence nucleon will give
non-vanishing contribution to the nuclear current which provides the
time-odd magnetic potentials.  The time-component of the vector
potential and the nuclear magnetic potential ( the spatial
components of the vector potential) in $^{17}$F are given in
Fig.\ref{fig1} as functions of $x$ for $y$=$z$=$1.29$ fm. Each
component of the nuclear magnetic potential has a peak value around
$1.0 \sim 1.5 $ MeV. Compared with the time-component of the vector
potential $V_0$, the nuclear magnetic potential is two order smaller
in magnitude. However, as the nucleon moves in a potential $\simeq
-50$ MeV given by the cancelation of the attractive scalar potential
$S(\simeq -400$ MeV) and the repulsive time-component of the vector
potential $V_0(\simeq +350$ MeV), the magnetic potential will play
important role in single-particle properties as will be shown later.
Furthermore, Fig.\ref{fig1} shows that the magnetic potential is
comparable in magnitude with the iso-vector vector potential
provided by the time-component of $\rho$ meson field.

\begin{figure}[ht]
 \centering
  \includegraphics[width=8cm]{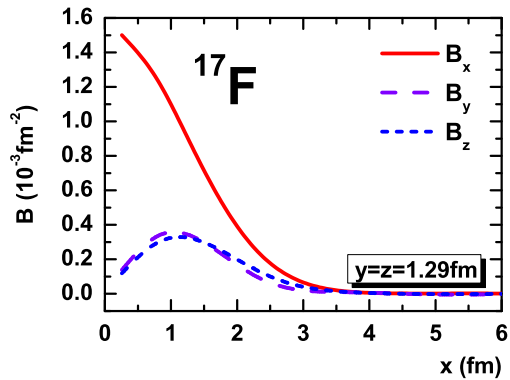}
   \caption{(color online) The nuclear magnetic field
    $\bB_n(=\nabla\times\bome)$
    versus $x$ axis for $y$=$z$=$1.29$ fm in $^{17}$F.}
   \label{fig2}
 \end{figure}

With the presence of the spatial-component of vector field, the
magnetic field $\bB_n$ can be defined as $\bB_n = \nabla \times
\bome$. The distribution of nuclear magnetic field $\bB_n$ versus
$x$ for $y$=$z$=$1.29$ fm in $^{17}$F are shown in Fig.\ref{fig2}.
It is found that $\bB_n$ changes rapidly with $x$ and its
$x$-component is larger than the other two components. The value
used in the estimation of the energy splitting between time
reversal conjugate states is around one third of the peak value
for the $x$-component of $\bB_n$ for $y$=$z$=$1.29$ fm in
$^{17}$F.
%
%

\begin{figure}[h!]
 \centering
 \includegraphics[width=8cm]{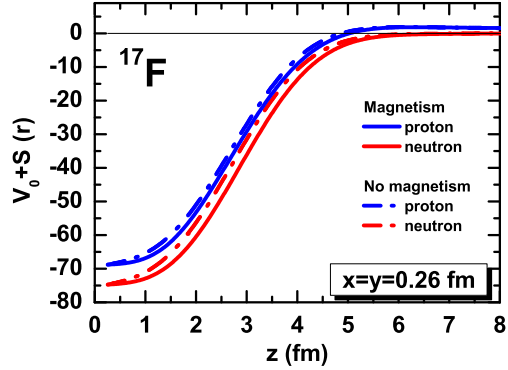}
 \caption{(color online) The potential $(V_0+S)(\br)$ for proton and neutron
 in $^{17}$F with (solid line) or without (dot-dashed line) the time-odd magnetic
 potentials self-consistently in triaxial RMF theory.}
 \label{fig3}
\end{figure}

Taking into account the time-odd potentials self-consistently will
lead to differences around $3$ MeV for $E_{\sigma}$ and $E_{\omega}$
in Eq.(\ref{Eq:MesonEngrgy}). Moreover, the nuclear magnetic
potential will lower the potential $(V_0+S)(\br)$ in $^{17}$F as
shown in Fig.\ref{fig3}. However, these differences will cancel with
each other. Therefore the time-odd magnetic potentials have small
influence on the binding energy, $rms$  radii and quadrupole moment
$Q_0$ for $^{17}$F. The binding energy per nucleon, $rms$ radii and
quadrupole moment for $^{17}$F in triaxial RMF theory with (without)
the time-odd magnetic potentials are respectively $7.600(7.589)$
MeV, $2.63 (2.63)$ fm and $9.78(9.86)$ fm$^2$. The deformation
parameters $\beta$ and $\gamma$ for $^{17}$F are respectively 0.08
and 57.14$^o$ after taking into account the time-odd magnetic
potentials self-consistently. The $\beta$ value is similar as the
previous axial deformed RMF calculation, i.e.,
$-0.086$\cite{Rutz98}.

%
%

%
%
%
%
 \subsection{Splitting of the time-reversal conjugate single-particle states}
 \label{sec:TC SPS}
%
%
%
{The violation of the time reversal invariance will result in the
core polarization and lead to the splitting of the the
time-reversal conjugate states. Moreover,} Because of deformation,
the single-particle angular momenta are no longer good quantum
numbers. The energies and expectation values of the angular
momentum for single-particle time reversal conjugate states
$\psi_{\unl j}$ and $\psi_{\ovl j}$ in $^{17}$F are respectively
given in Table \ref{Table1}. The expectation values of the orbital
angular momentum for the large ($\langle L \rangle$) and small
component ($\langle L^\prime\rangle$) of the single-particle state
are close to their spherical values due to the small $\beta$. As
$\gamma$ is close to 60$^o$, $L_x+S_x$ is an approximate good
quantum number and close to its axial deformed value, from which
we can determine the approximate quantum number $ljm$ for the
single-particle states as given in Table \ref{Table1} in this
case. One also has $\langle L^\prime\rangle \simeq \langle
L\rangle\pm 1$ for the state $J=L\pm\frac{1}{2}$. As expected from
the estimation in Sec.\ref{sec:sec1}, the energy splitting for
single-particle time reversal conjugate states in $^{17}$F range
from $0.04$ to $0.7$ MeV and larger splitting occurs for states
with larger expectation value $\langle L_x+S_x \rangle$.

%
 \begin{table*}[ht!]
 \begin{center}
   \tabcolsep=3pt
    \caption{The energies and expectation values of the angular
momentum for single-particle time reversal conjugate states
$\psi_{\unl j}$ and $\psi_{\ovl j}$ in $^{17}$F, where $L$ and
$L^\prime$ are the orbital angular momenta of large and small
components respectively, $L_x$ and $S_x$ are the $x$-component of
the orbital angular momenta and spin, and $\Delta \varepsilon
=\varepsilon_{\unl j} - \varepsilon_{\ovl j}$.}
    \begin{tabular}{|c|ccccc|ccccc|c|}
    \hline
    \multicolumn{12}{|c|}{\bfseries proton}\\
\hline
 $l_{j|m|}$&  \multicolumn{5}{c|}{\bfseries $\psi_{\unl j}$}&\multicolumn{5}{c|}{\bfseries $\psi_{\ovl j}$}&\\
    \hline
       &  $\varepsilon$ (MeV) &  $L$ $(\hbar)$ & $L^\prime$ $(\hbar)$ &  $L_x$  $(\hbar)$&  $S_x$  $(\hbar)$
       & $\varepsilon$ (MeV) &  $ L $ $(\hbar)$ & $ L^\prime$ $(\hbar)$ & $L_x$ $(\hbar)$&  $S_x$  $(\hbar)$& $\Delta \varepsilon$ (MeV)\\
\cline{2-12}
$s_{\frac{1}{2}\frac{1}{2}}$  & -37.55 &  0.01  &  1.01   &  0.00  &  -0.50 &  -37.92  & 0.01  & 1.01    & 0.00    & 0.50 & 0.37\\
$p_{\frac{3}{2}\frac{3}{2}}$  & -17.93 &  1.01  &  1.93   &  -1.00 &  -0.50 &  -18.60  & 1.01  & 1.92    & 1.00    &0.50 &0.67\\
$p_{\frac{3}{2}\frac{1}{2}}$  &  -17.92  & 1.01  & 1.97 & -0.41& -0.10      & -17.88   &  1.01 &  1.97   &  0.40   &  0.10 & -0.04  \\
$p_{\frac{1}{2}\frac{1}{2}}$   &  -11.60  & 1.03  & 0.28    & -0.59   & 0.10& -11.83 &  1.03  &  0.28   &  0.60  &  -0.10& 0.23 \\
$d_{\frac{5}{2}\frac{5}{2}}$  & -1.14  &  1.98  &  2.85   &  -1.98 &  -0.50 &  -1.83   & 1.92  & 2.83    & 2.00    & 0.50 & 0.69 \\
$d_{\frac{5}{2}\frac{3}{2}}$  &  -1.10 & 2.02   & 2.91    & -0.95  & -0.20  & -1.27  &  2.03  &  2.94   &  1.22  &   0.25 &0.17 \\
$d_{\frac{5}{2}\frac{1}{2}}$  & -1.01  &  2.01  &  2.93   &  -0.39  &  -0.10 &  -1.08   & 2.02  & 2.98   & 0.12    & 0.03 & 0.07 \\
         \hline
    \multicolumn{12}{|c|}{\bfseries neutron}\\
\hline
  $l_{j|m|}$      &  \multicolumn{5}{c|}{\bfseries $\psi_{\unl j}$}&\multicolumn{5}{c|}{\bfseries $\psi_{\ovl j}$}&\\
    \hline
    &  $\varepsilon$ (MeV) &  $L$ $(\hbar)$ & $L^\prime$ $(\hbar)$ &  $L_x$  $(\hbar)$&  $S_x$  $(\hbar)$
       & $\varepsilon$ (MeV) &  $ L $ $(\hbar)$ & $ L^\prime$ $(\hbar)$ & $L_x$ $(\hbar)$&  $S_x$  $(\hbar)$& $\Delta \varepsilon$ (MeV)\\
\cline{2-12}
$s_{\frac{1}{2}\frac{1}{2}}$ & -43.72 &  0.01  &  1.01   &  0.00  &  -0.50  &  -44.10  & 0.01  & 1.01 &  0.00  & 0.50 &0.38\\
$p_{\frac{3}{2}\frac{3}{2}}$ & -24.26 &  1.01  &  1.93   &  -1.00 &  -0.50 &  -24.96  & 1.01  & 1.93 &  1.00 & 0.50 &0.70\\
$p_{\frac{3}{2}\frac{1}{2}}$  &  -23.73  & 1.01  & 1.95 &  -0.48  & -0.02& -23.69 &  1.01  &  1.96   &  0.47  &  0.03  &-0.04 \\
$p_{\frac{1}{2}\frac{1}{2}}$  &  -17.40  & 1.02  & 0.33 &  -0.52  & 0.02 & -17.65 &  1.02  &  0.33   &  0.53  &  -0.03 & 0.25\\
$d_{\frac{5}{2}\frac{5}{2}}$ & -6.93 &  1.93  &  2.81   &  -2.00  &  -0.50  &  -7.69  & 1.93  & 2.82 &  2.00  & 0.50&0.76\\
$d_{\frac{5}{2}\frac{3}{2}}$  &  -6.35   & 2.00  & 2.91 & -1.23 & -0.21 & -6.54  &  2.00  &  2.93   &  1.27  &   0.22 & 0.19  \\
$d_{\frac{5}{2}\frac{1}{2}}$ & -6.15  &  1.97  &  2.90   &  -0.38  &  -0.11 &  -6.22   & 1.98  & 2.91 & 0.34 & 0.10 & 0.07 \\
    \hline
    \end{tabular}
    \label{Table1}
   \end{center}
  \end{table*}
%
%

\begin{figure}[ht]
 \centering
 \includegraphics[width=8cm]{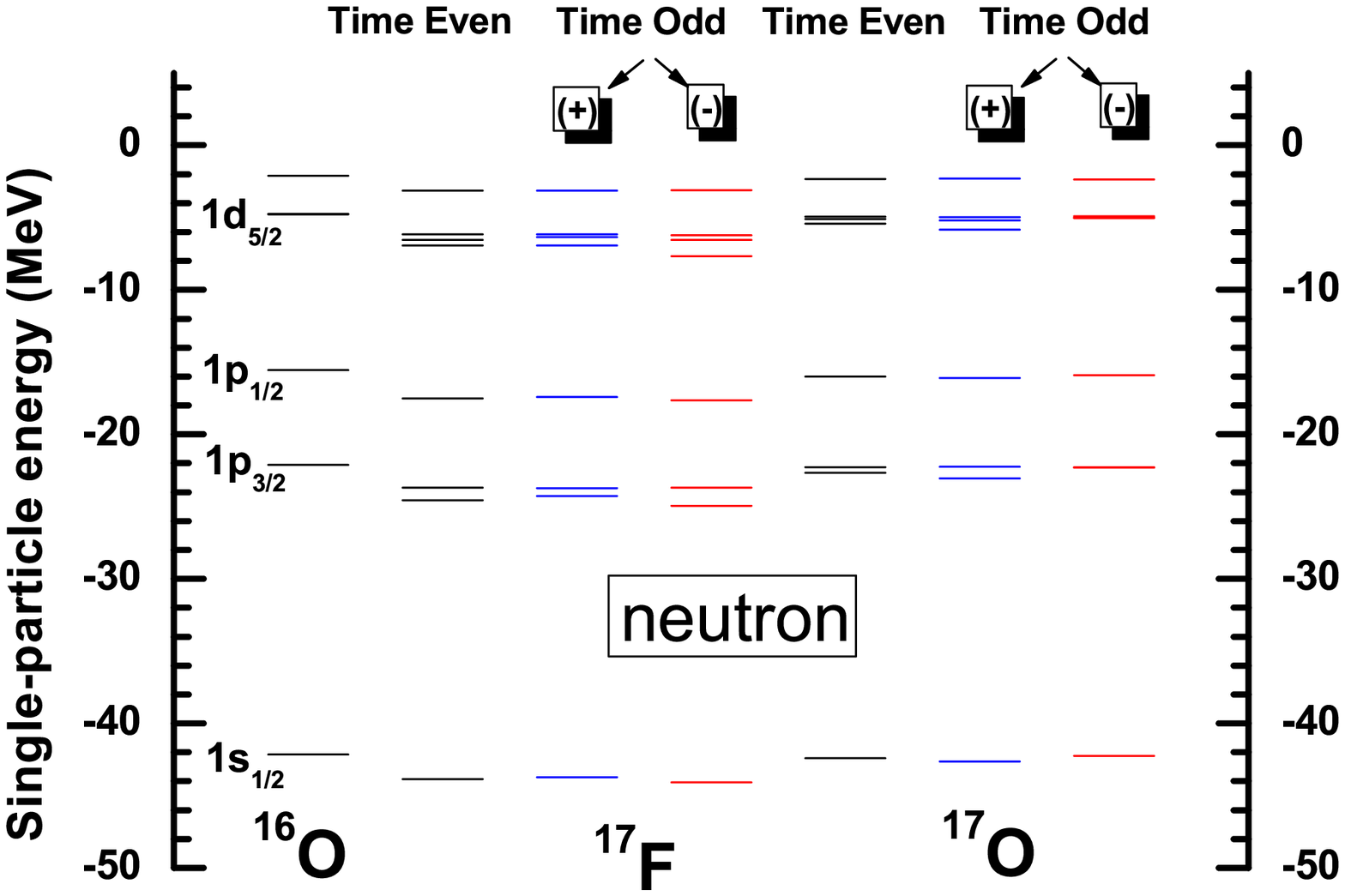}
 \includegraphics[width=8cm]{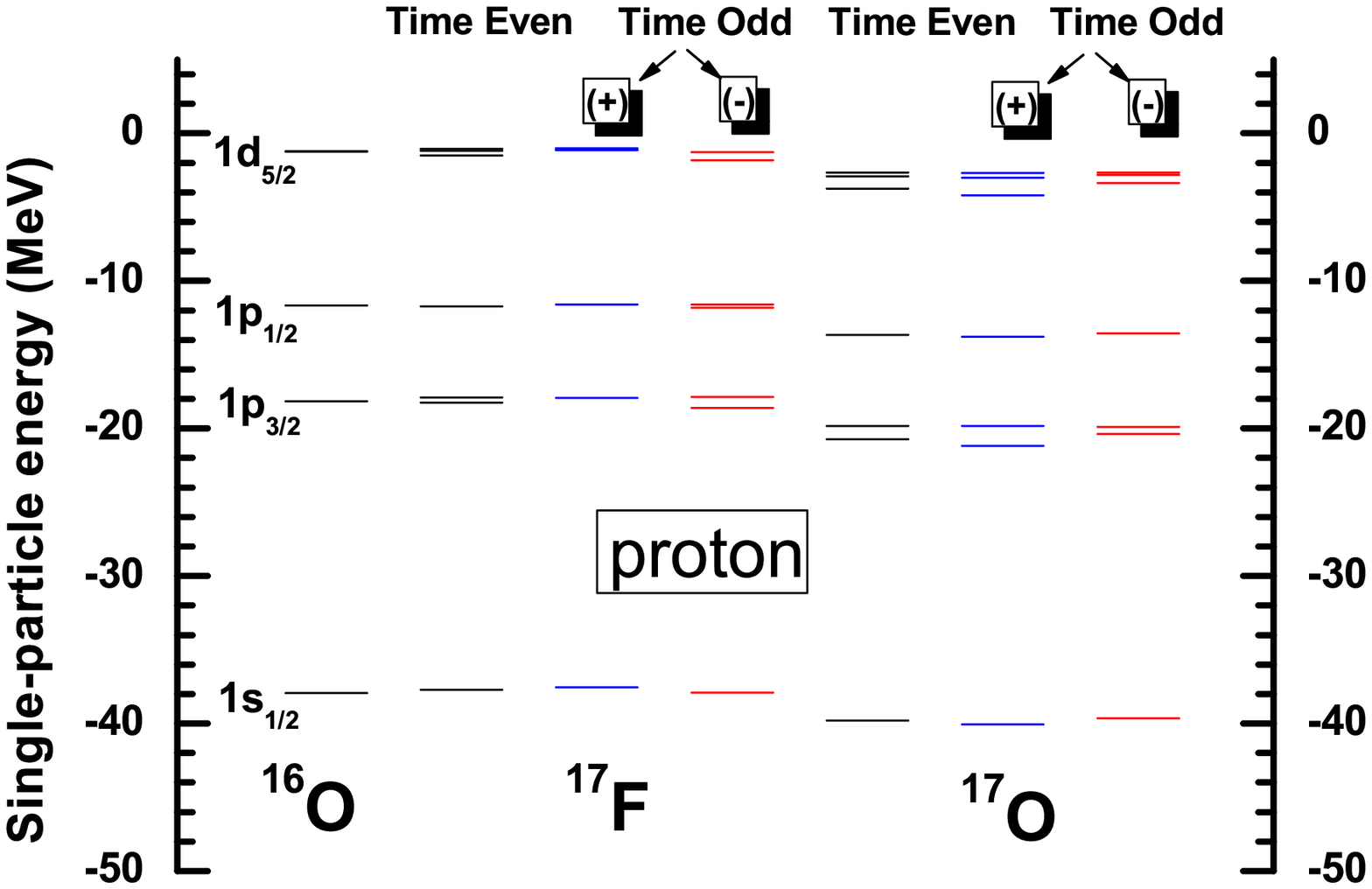}
\caption{(color online) The proton and neutron single-particle
levels for time reversal conjugate states in $^{17}$O and $^{17}$F.
Those in $^{16}$O as well as in $^{17}$O and $^{17}$F without
time-odd potentials are also given as a reference. The  levels with
negative angular momentum projection are given in blue (the 3rd and
6th columns) and the levels with positive angular momentum
projection are given in red (the 4th and 7th columns) . }
 \label{fig4}
\end{figure}

In order to investigate the difference between the energy splitting
due to the violation of time reversal invariance from valence
neutron and proton, the proton and neutron single-particle levels
for time reversal conjugate states in $^{17}$O and $^{17}$F are
shown in Fig.\ref{fig4}. As a reference, those in $^{16}$O as well
as in $^{17}$O and $^{17}$F with time-even potentials only are also
given since $^{17}$O or $^{17}$F is nothing but one neutron or
proton on top of the $^{16}$O core.

There are the following features: 1) The valence proton (neutron)
in $^{17}$F ($^{17}$O) will lower the neutron (proton) single
particle levels more noticeably than  the corresponding  proton
(neutron) single particle levels; 2) The energy splitting due to
the violation of time reversal invariance from valence proton
(neutron) in $^{17}$F ($^{17}$O) is distinguishable in
Fig.\ref{fig4} which denotes the necessity to take into account
the time-odd magnetic potentials. In Ref. \cite{Rutz98}, the
influence of the time-odd magnetic potentials on single-particle
energies and the related quantities such as single-particle
separate energy has been investigated.


%
%

\begin{figure}[ht]
 \centering
 \includegraphics[width=8cm]{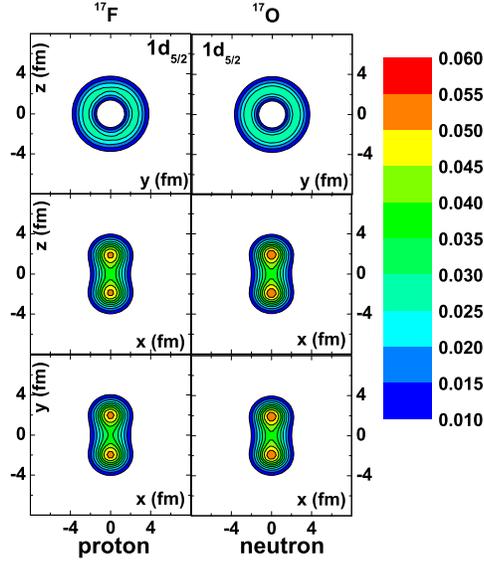}
  \caption{(color online)The density distribution of the unpaired valence nucleon in $yz$,
  $zx$, and $xy$ plane for $^{17}$F and $^{17}$O while the other axis $x$, $y$, and $z$
  is integrated respectively.}
 \label{fig5}
\end{figure}

\begin{figure}[ht]
 \centering
 \includegraphics[width=8cm]{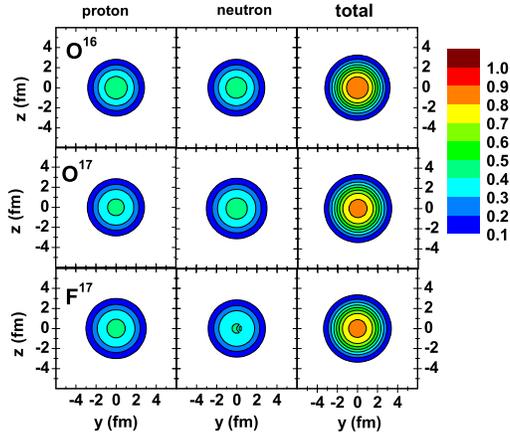}
  \caption{(color online) The proton, neutron and matter density distribution
  in $yz$, $zx$, and $xy$ plane for $^{16}$O, $^{17}$O, and $^{17}$F, while
  the other axis $x$, $y$, and $z$
  is integrated respectively. }
 \label{fig6}
\end{figure}

The density distribution of the unpaired valence nucleon for
$^{17}$O and $^{17}$F as well as the proton, neutron and matter
for $^{16}$O, $^{17}$O, and $^{17}$F in $yz$, $zx$, and $xy$ plane
are plotted in Figs.\ref{fig5} and \ref{fig6} respectively, in
which the other axis $x$, $y$, and $z$ is integrated respectively.
It is interesting to observe that although there are noticeable
differences in the energy of the unpaired valence nucleon,  the
density distributions of the valence proton in $^{17}$F and
neutron in $^{17}$O are quite similar to each other and they are
corresponding to $1d_{5/2}$ orbits. In Fig.\ref{fig6}, the density
distributions for proton, neutron and matter in $^{17}$O and
$^{17}$F as well as $^{16}$O are shown.

%

%
%

%
\begin{figure}[ht]
 \centering
 \includegraphics[width=6cm]{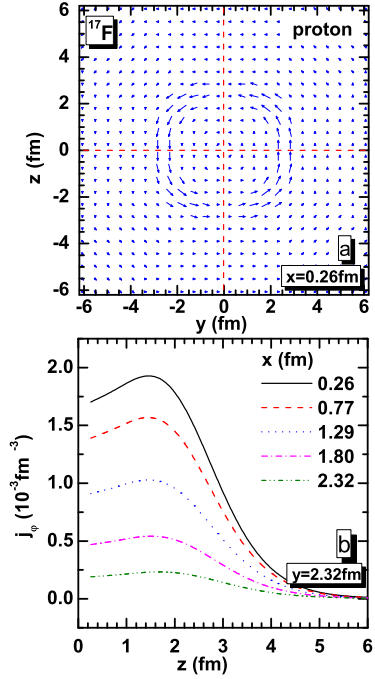}
 \caption{(color online) Upper: The Dirac current for proton in $yz$
 plane at x=0.26 fm in $^{17}$F. The direction and length of the arrows
 respectively represent the orientation and magnitude
 of current. Lower: The azimuthal component of Dirac current
 for proton as a function of $z$ with y=2.32 fm and
 x=0.26, 0.77, 1.29, 1.80, 2.32 fm respectively in $^{17}$F.
 }
 \label{fig7}
\end{figure}
%
%
%
 \subsection{Current and nuclear magnetic moments}
 \label{sec:CurrentMoment}
%
%
%
 Nuclear magnetic moment is related to the effective electromagnetic
 current operator~\cite{Furnstahl89,Serot81}
 \beq
  \label{EMcurrent}
   \hat J^\mu(x)=Q\bar\psi(x)\gamma^\mu\psi(x)+\frac{\kappa}{2M}\partial_\nu
   [\bar\psi(x)\sigma^{\mu\nu}\psi(x)],
 \eeq
where the field operators are in the Heisenberg representation with
$Q\equiv\frac{1}{2}(1-\tau_3)$, $M$ the nucleon mass,
$\sigma^{\mu\nu}=\frac{i}{2}[\gamma^\mu,\gamma^\nu]$, and $\kappa$
the free anomalous gyromagnetic ratio of the nucleon:
$\kappa^p=1.793$ and $\kappa^n=-1.913$. The spatial-component of the
current operator is given by,
 \beq
  \label{current}
  \bj(\br)=Q\psi^\dagger(\br)\balp\psi(\br)+\frac{\kappa}{2M}\nabla\times
  [\psi^\dagger(\br)\beta\bSig\psi(\br)],
 \eeq
where the first term gives the Dirac current
$\bj_D(\br)=\psi^\dagger(\br)\balp\psi(\br)$, which can be
decomposed into a orbital (convection) current and a spin current
according to Gordon identity, and the second term in
Eq.~(\ref{current}) is the so-called anomalous current.

The proton Dirac current in $yz$ plane at x=0.26 fm in $^{17}$F is
given in the upper panel in Fig.\ref{fig7} in which the direction
and length of the arrows respectively represent the orientation
and magnitude of current. It is clearly seen that the Dirac
current peaks at the nuclear surface. In order to investigate how
the Dirac current change with $x$, the azimuthal component of
Dirac current for proton is given for $x=0.26, 0.77, 1.29, 1.80,
2.32$ fm respectively as a function of $z$ with $y=2.32$ fm in
$^{17}$F. It is found that the Dirac current decreases with $x$.
Compared with the density distribution for the valence nucleon in
Fig.~\ref{fig5}, it is obvious that the current is mainly from the
valence nucleon and moves around the nuclear surface in $yz$
plane. In order to see the core polarization effect, the nuclear
magnetic moment in the following is more suitable than the
current.

\begin{figure}[ht!]
 \centering
 \includegraphics[width=6cm]{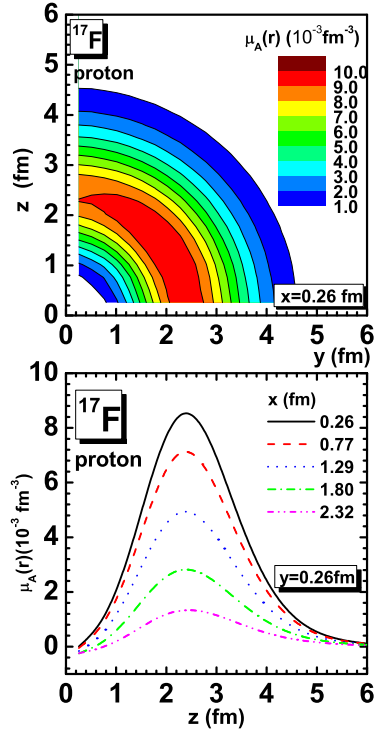}
 \caption{(color online)
 Upper: The anomalous magnetic moment
 density $\mu_A(\br)$ for proton in $yz$
 plane at x=0.26 fm in $^{17}$F.
 Lower: The anomalous magnetic moment
 density $\mu_A(\br)$ for proton as a function of $z$ with y=0.26 fm and
 x=0.26, 0.77, 1.29, 1.80, 2.32 fm respectively in $^{17}$F. }
 \label{fig8}
\end{figure}

In the mean field approximation, the nuclear magnetic moment can
be calculated from the expectation value of the current operator
in Eq.~(\ref{current}) in ground state,
 \beqn
 \label{NMM}
   \bmu &=&2M \int d^3r
   \frac{1}{2}[\br \times  \langle g.s. | \bj(\br) | g.s. \rangle]\nonumber\\
   &=&\sum\limits_{i=1}^A\int d^3r
   \left[\frac{Mc^2}{\hbar c}Q\psi^\dagger_i(\br)\br\times\balp\psi_i(\br)
   +\kappa\psi^\dagger_i(\br)\beta\bSig\psi_i(\br)\right].\nonumber\\
 \eeqn
Here, $\hbar$ and $c$ are added in order to give the magnetic moment
in units of nuclear magneton. The magnetic moment can be divided as
the Dirac magnetic moment,
 \beq
 \label{DMM}
  \bmu_D=\sum\limits_{i=1}^A\frac{Mc^2}{\hbar c}\int d^3r
  Q\psi^\dagger_i(\br)\br\times\balp\psi_i(\br),
  \eeq
and the anomalous magnetic moment,
 \beq
 \label{AMM}
  \bmu_A= \kappa \int d^3r \mu_A(\br) =\sum \limits_{i=1}^A \kappa \int d^3r \psi^\dagger_i(\br)\beta\bSig\psi_i(\br).
  \eeq
  The anomalous magnetic moment (proton)
 density $\mu_A(\br)$ distribution in x=0.26 fm plane
 and the distribution of $\mu_A(\br)$ (proton) along $z$ in y=0.26 fm plane,
 with x=0.26, 0.77, 1.29, 1.80, 2.32 fm respectively in $^{17}$F
 are plotted in Fig.\ref{fig8}. It is found that the anomalous
 magnetic moment density distribution has the similar character to
 the Dirac current since they are both mainly dependent on the density distribution
 of the unpaired valence proton given by Fig.\ref{fig5}.
%


%

\begin{table*}[ht!]
  \begin{center}
  \tabcolsep=14pt
  \caption{The magnetic moments of light nuclei near double-closed shells
  in units of nuclear magneton $\mu_N$. The spherical and axial deformed
  calculations with NL1 are taken from Ref.\cite{Hofmann88}. }
   \begin{tabular}{ccccccccc}
    \hline
     \hline
              $\mu$  & $^{15}$O&$^{17}$O&$^{39}$Ca&$^{41}$Ca&$^{15}$N&$^{17}$F&$^{39}$K&$^{41}$Sc \\
      \hline
          Exp.    & 0.72   & -1.89  & 1.02   & -1.60         & -0.28    & 4.72   & 0.39     & 5.43 \\
          Tri.RMF   & 0.57   & -2.00   & 0.98  & -2.13       &  -0.19   & 4.893  & 0.37   & 6.04  \\
          Axi.RMF   & 0.65   & -2.03  & 0.96   & -2.13        & -0.29    & 4.99   & 0.33   & 6.07\\
          Sph.RMF   & 0.66   & -1.91  & 1.17   & -1.91        & -0.03    & 5.05   & 0.72   & 6.32\\
          Schmidt   & 0.64   & -1.91  & 1.15   & -1.91        & -0.26    & 4.79   & 0.12   & 5.79 \\
        \hline\hline
  \end{tabular}
   \label{table2}
   \end{center}
 \end{table*}


The magnetic moments in Eq.(\ref{NMM}) as well as the Dirac and
anomalous magnetic moments in Eqs. (\ref{DMM}) and (\ref{AMM}) for
light nuclei near double-closed shells with A=15, 17, 39, 41 are
respectively given in Tables \ref{table2} and \ref{table3}.

\begin{table*}[ht!]
  \begin{center}
  \tabcolsep=14pt
  \caption{The Dirac and anomalous magnetic moments ( $\mu_D$ and
  $\mu_A$ ) of light nuclei
   near closed shells in units of $\mu_{\rm N}$. The spherical and axial deformed
  calculations with NL1 are taken from Ref.\cite{Hofmann88}.}
   \begin{tabular}{ccccccccc}
    \hline
     \hline
              $\mu_D$  & $^{15}$O&$^{17}$O&$^{39}$Ca&$^{41}$Ca&$^{15}$N&$^{17}$F&$^{39}$K&$^{41}$Sc \\
      \hline
          Tri.RMF    & -0.11 & -0.13  & -0.16   & -0.28       & 0.46     & 3.15   & -0.64    & 4.31   \\
          Axi.RMF    & -0.13 & -0.13  & -0.30   & -0.22       & 0.44     & 3.21   & 1.50     & 4.29  \\
          Sph.RMF    & -     &  -     &   -     &  -          & 0.59     & 3.26   & 1.81     & 4.54  \\
          Schmidt    & -     &  -     &   -     &  -          & 0.33     & 3.00   & 1.20     & 4.00  \\
 \hline
           $\mu_A$  & $^{15}$O&$^{17}$O&$^{39}$Ca&$^{41}$Ca&$^{15}$N&$^{17}$F&$^{39}$K&$^{41}$Sc \\
 \hline
           Tri.RMF    & 0.68 &  -1.87  &  1.13  & -1.85        & -0.64    & 1.75  &  1.00     &1.74   \\
           Axi.RMF    & 0.78 &  -1.90  &  1.26  & -1.87        & -0.73    & 1.78  & -1.17     & 1.79 \\
           Sph.RMF    & 0.66 &  -1.91  &  1.17  & -1.91        & -0.62    & 1.79  & -1.09     & 1.79  \\
          Schmidt     & 0.64 &  -1.91  &  1.15  & -1.91        & -0.60    & 1.79  &  -1.08    & 1.79  \\
 \hline\hline
  \end{tabular}
   \label{table3}
   \end{center}
 \end{table*}
 \begin{table*}[ht!]
  \begin{center}
    \caption{Iso-scalar magnetic moments $\mu(A)=[\mu(Z,N)+\mu(Z+1,N-1)]/2$
    of light nuclei near closed shells in units of $\mu_{\rm N}$.}
  \begin{tabular}{ccccccccc}
    \hline
     A   &   Orbit         & EXP.    & Tri.RMF   & Axi.RMF\cite{Hofmann88}  & Sph.RMF\cite{Hofmann88}  & Schmidt  \\
      \hline
     15  &  $1p_{1/2}$     &  0.22   &   0.19    &  0.18                    &   0.32                   &  0.19     \\
     17  &  $1d_{5/2}$     &  1.41   &   1.45    &  1.48                    &   1.57                   &  1.44     \\
     39  &  $1d_{3/2}$     &  0.71   &   0.67    &  0.64                    &   0.94                   &  0.64     \\
     41  &  $1f_{7/2}$     &  1.92   &   1.96    &  1.97                    &   2.21                   &  1.94     \\
     \hline
   \end{tabular}
    \label{table4}
   \end{center}
  \end{table*}
%

 \begin{table}[ht!]
  \begin{center}
   \caption{Iso-vector magnetic moments $\mu(A)=[\mu(Z,N)-\mu(Z+1,N-1)]/2$
   of light nuclei near closed shells in units of $\mu_{\rm N}$.}
    \begin{tabular}{ccccccc}
     \hline\hline
      A   &  Exp.   & Tri.RMF     & Axi.RMF\cite{Hofmann88} &  Sph.RMF\cite{Hofmann88} & Schmidt  \\
   \hline
      15  &  0.501  &0.376        &0.470                    &            0.345         &  0.451   \\
      17  &  -3.303 &-3.446       &-3.510                   &            -3.480        &  -3.353  \\
      39  &  0.312  &0.305        &0.315                    &             0.225        &  0.512   \\
      41  &  -3.513 &-4.086       &-4.10                    &            -4.115        &  -3.853  \\
     \hline\hline
   \end{tabular}
    \label{table5}
   \end{center}
  \end{table}

As shown in Table \ref{table2}, the magnetic moments given by
time-odd axial deformed RMF and triaxial deformed RMF calculations
reproduce the data well in most cases and are close to each other
due to the small $\gamma$ deformation for the nuclei investigated
here. The relative large difference for $^{15}$N in Table
\ref{table2} between the axial and triaxial RMF calculations is
mainly from the anomalous part and it is due to the inclusion of
$\gamma$ degree of freedom.

The anomalous magnetic moments given by the triaxial RMF
calculations are close to those of the spherical ones. While the
difference in the total magnetic moment between the spherical and
triaxial calculations is mainly from the Dirac magnetic moments
due to the core polarization. Without core polarization, the Dirac
magnetic moment of neutron particle (hole) should vanish. However,
as seen in Table \ref{table3},  the Dirac magnetic moment for
$^{15}$O, $^{17}$O, $^{39}$Ca, and $^{41}$Ca contribute around
10\% to the total nuclear magnetic moment which is the effect of
the core polarization. The core polarization can be taken into
account self-consistently in the axial and triaxial calculations
but not in the spherical one.

Although the Dirac and anomalous magnetic moments for $^{39}$K
given by the triaxial RMF calculation have different signs with
the other calculations, the triaxial RMF calculation better
reproduce the experimental magnetic moment. The triaxial deformed
RMF calculation result of $^{39}$K gives the deformation
parameters $\beta=0.038$ and $\gamma=58.66^0$.

The iso-scalar and iso-vector magnetic moments from the spherical,
axial and triaxial deformed RMF calculations for light nuclei near
the closed shells are given in Tables \ref{table4} and
\ref{table5} in comparison with the data and Schmidt values. As
mentioned before, the iso-scalar magnetic moments in deformed RMF
calculations are close to the Schmidt values if the core
polarization effect has been taken into account self-consistently.
The spherical RMF calculations, however, have discrepancies around
$10 \sim 30 \%$ with the data. For the iso-vector magnetic
moments, the axial and triaxial deformed RMF calculations agree
better with the data than the spherical RMF calculations and the
Schmidt values, which demonstrates the importance of the core
polarization. In fact, the agreement with the experimental
iso-scalar and iso-vector magnetic moments can be improved also in
the spherical RMF calculations after considering the core
polarization by either the Landau-Migdal approach \cite{McNeil86}
or the configuration mixing \cite{Nedjadi89}.

 \section{Summary}
 \label{sec:sec4}

The time-odd triaxial relativistic mean field approach has been
developed and applied to the investigation of the ground-state
properties of light odd-mass nuclei near double-closed shells. The
splitting due to the violation of the time reversal invariance in
the single-particle energy has been estimated by reducing the Dirac
equation with the time-odd nuclear magnetic potential to Schrodinger
equation.

The Dirac equation with time-odd nuclear magnetic potential for
nucleon and Klein-Gordon equations for meson fields have been solved
self-consistently by expansion on the three dimensional harmonic
oscillator basis with PK1. Taking $^{17}$F as an example, the
ground-state properties including the binding energy, deformation
$\beta$ and $\gamma$, the single-particle energy and the splitting
of the time reversal conjugate states, density and current
distribution, and magnetic moments, etc., have been examined. It is
found that although the non-vanishing spatial-component of the
$\omega$ field due to the violation of time reversal symmetry in
odd-mass nuclei has small influence on the binding energy,
root-mean-square radii and quadrupole moment, it will create a
magnetic potential, change the nuclear wave function and result in
the core polarization effect which plays important role on the
single-particle properties and the magnetic moment. The nuclear
magnetic moments for the light double-closed shells nuclei plus or
minus one nucleon, including the Dirac, anomalous, iso-scalar, and
iso-vector magnetic moments, have been investigated and good
agreements with the data have been achieved.

With the time-odd triaxial relativistic mean field approach
developed here, it can be expected to describe well not only the
light double-closed shells nuclei plus or minus one nucleon, but
also the nuclei in the $\gamma$-soft mass region. Investigation
along this line is in progress.

\begin{acknowledgments}

This work is partly supported by the National Natural Science
Foundation of China under Grant No. 10435010, 10575083 and 10221003.

\end{acknowledgments}


\begin{appendix}

 \section{The basis for solving the Dirac equation}
  \label{AppendixA}

Using the eqs. (\ref{WaveFunction}) and (\ref{combination}), the
Dirac equation (\ref{DiracEq}) with time-odd magnetic potentials can
be written in matrix form,
 \beqn
 \label{matrix}
  \left(
   \begin{array}{cccc}
    A_{\alpha'\alpha} & {\cal A_{\alpha'\ovl\alpha}}&
   {\cal B_{\alpha'\alpha}} & B_{\alpha'\ovl\alpha}\\
   {\cal A_{\ovl\alpha'\alpha}} & A_{\ovl\alpha'\ovl\alpha} &
    B_{\ovl\alpha'\alpha} & {\cal B_{\ovl\alpha'\ovl\alpha}}\\
  {-\cal B_{\beta'\alpha}} & -B_{\beta'\ovl\alpha} &
    C_{\beta'\alpha} &{\cal C_{\beta'\ovl\alpha}}\\
   -B_{\ovl\beta'\alpha} & {-\cal B_{\ovl\beta'\ovl\alpha}} &
   {\cal C_{\ovl\beta'\alpha}} & C_{\ovl\beta'\ovl\alpha}\\
  \end{array}
   \right)
  \left(
     \begin{array}{c}
     f_{\alpha}\\
     f_{\ovl\alpha}\\
     g_{\alpha}\\
     g_{\ovl\alpha}
   \end{array}
   \right)=\varepsilon\left(
     \begin{array}{c}
     f_{\alpha^\prime}\\
     f_{\ovl\alpha^\prime}\\
     g_{\beta^\prime}\\
     g_{\ovl\beta^\prime}
   \end{array}
   \right).\nonumber\\
  \eeqn
 If the mirror symmetries with respect to three planes ($xy$), ($yz$) and ($xz$)
 are assumed for all scalar potentials and vector potentials \cite{Konig93},
 it can be shown that the matrix elements in script letters will vanish.
 For example, the invariance of the matrix element ${\cal A}_{\alpha'\ovl\alpha}$,
 \begin{widetext}
 \beqn
 {\cal A}_{\alpha'\ovl\alpha}&=&\frac{1+(-1)^{n'_x+\tilde{n}_x+1}}{2}(-1)^{\tilde n_x+\tilde n_y+1}\cdot(-1)^{\frac{\tilde n_y-n'_y}{2}}\nonumber\\
 && \cdot\int d^3r\phi_{n'_x}(x)\phi_{n'_y}(y)\phi_{n'_z}(z)
  (V_0+M^*)\phi_{\tilde n_x}(x)\phi_{\tilde n_y}(y)\phi_{\tilde n_z}(z)
 \eeqn
 \end{widetext}
 under the operation $x\rightarrow -x$ will
 leads to ${\cal A}_{\alpha'\tilde\alpha}=0$. After similar
 operation for $y$ and $z$, the matrix in Eq.(\ref{matrix})
 can be written as,
  \beqn
   \left(
   \begin{array}{cccc}
    A_{\alpha'\alpha}     & B_{\alpha'\ovl\alpha}   &  0 & 0\\
    -B_{\ovl\beta'\alpha} &C_{\ovl\beta'\ovl\alpha} &  0 & 0\\
   0& 0 & A_{\ovl\alpha'\ovl\alpha} & B_{\ovl\alpha'\alpha} \\
   0 & 0 & -B_{\beta'\ovl\alpha} &  C_{\beta'\alpha} \\
  \end{array}
   \right)
  \left(
     \begin{array}{c}
     f_{\alpha}\\
     g_{\ovl\alpha}\\
     f_{\ovl\alpha}\\
    g_{\alpha}\\
   \end{array}
   \right)=\varepsilon\left(
     \begin{array}{c}
     f_{\alpha^\prime}\\
    g_{\ovl\beta^\prime}\\
     f_{\ovl\alpha^\prime}\\
     g_{\beta^\prime}\\
   \end{array}
   \right),\nonumber\\
  \eeqn
 which indicates that the space $\{\psi_i\}$ can be reduced into two
 subspace $\{\psi_{\unl j}\}$ and $\{\psi_{\ovl j}\}$ in Eq.(\ref{TimeEven}).

\end{appendix}

\end{document}